\documentclass[10pt,twocolumn]{article}
\usepackage{Nsc2016}

\usepackage{algpseudocode,algorithm}
\usepackage{graphicx} 
\usepackage{subfig} 
\usepackage{icomma} 
\usepackage{amsmath,amssymb,amsfonts} 

\usepackage{graphicx} 
\usepackage{psfrag} 
\usepackage{float} 
\usepackage[framed]{mcode}

\usepackage[numbers,sort&compress]{natbib} 
\usepackage{url} 
\usepackage[numbib,nottoc]{tocbibind} 

\title{\uppercase{SIMULATION OF CHUA'S CIRCUIT BY MEANS OF INTERVAL ANALYSIS}}

\author[1]{\underline{Silva, Melanie Rodrigues}}
\author[2]{Nepomuceno, E. G.}
\author[3]{Amaral, G. F. V.}
\author[4]{Silva, V. V. R.}

\affil[1]{Modelling and Control Research Group - GCOM, Graduate Program in Electrical Engineering - PPGEL, Federal University of S\~{a}o Jo\~{a}o del-Rei - UFSJ, S\~{a}o Jo\~{a}o del-Rei, Brazil, me\_rodrigues\_silva@hotmail.com}
\affil[2]{Modelling and Control Research Group - GCOM, Department of Electrical Engineering - DEPEL, Federal University of S\~{a}o Jo\~{a}o del-Rei - UFSJ, S\~{a}o Jo\~{a}o del-Rei, Brazil, nepomuceno@ufsj.edu.br}
\affil[3]{Modelling and Control Research Group - GCOM, Department of Electrical Engineering - DEPEL, Federal University of S\~{a}o Jo\~{a}o del-Rei - UFSJ, S\~{a}o Jo\~{a}o del-Rei, Brazil, amaral@ufsj.edu.br}
\affil[4]{Department of Electrical Engineering - DEPEL, Federal University of S\~{a}o Jo\~{a}o del-Rei - UFSJ, S\~{a}o Jo\~{a}o del-Rei, Brazil, vvrsilva@ufsj.edu.br}

\begin{document}

\maketitle
\maketitle

\Abstract{The Chua's circuit is a paradigm for nonlinear scientific studies. It is usually simulated by means of numerical methods under IEEE 754-2008 standard. Although the error propagation problem is well known, little attention has been given to the relationship between this error and inequalities presented in Chua's circuit model. Taking the average of round mode towards $+\infty$ and $-\infty$, we showed a qualitative change on the dynamics of Chua's circuit.} 


\section{INTRODUCTION}

We live in a complex world \citep{Monteiro} where the majority of dynamical systems are nonlinear. The choice of non-linear models brings with it an inevitable increase in the complexity \citep{aguirre}. One of the systems most widely used as a model to study the nonlinear dynamics and chaos is the Chua's circuit \cite{kennedy92}.

The Chua's circuit, developed by Leon O. Chua in 1984, is known to exhibit chaotic behaviour similar to the system proposed by Lorenz \citep{chua94}. It is a simple electronic network Which exhibits a variety of phenomena, such as strange attractors and bifurcations. The circuit Consists of two capacitors, an inductor, a linear resistor, and a nonlinear resistor \citep{kennedy92}.

The numerical computation uses floating point arithmetic in most computers under the IEEE 7542-2008 standard \citep{Gold,ieee}. This standard, in some sense, is a systematic approximation of the real arithmetic and it is represented by a finite subset of the real numbers. As a result, some properties of the arithmetic of real numbers are not guaranteed for the floating-point \citep{ieee}. So, it is necessary attention to the limitations of computers with respect to scientific computing.

Given the arithmetic floating point is an approximation of the real numbers, small errors are generated during the process and simulations \citep{Nepomuceno2014}. These errors in the eyes of many users and are made subjectively considered unnoticeable, but can strongly influence the results.


This article focuses on the computational numerical round modes. The inequalities presented the Chua's circuit model  are addressed in parallel with interval analysis \citep{moore1979methods} of the circuit.  Taking the average of round mode towards $+\infty$ and $-\infty$, we showed a qualitative change on the dynamics of Chua's circuit.

\section{OBJECTIVES}

The purpose of this article is to show the influence of computer round modes applied to Chua's circuit. Using interval arithmetic circuit in the simulation mode to a careful analysis of these results and those originating traditional computational rounding, that is, according to the rules of IEEE standard 754-2008 floating-point arithmetic. Since the circuit is chaotic behaviour, the influence of error propagation in the system dynamics is also the target of the study.

\section{METHODS} 

\begin{figure}[!hbt]
			\centering
			\subfloat[]
			{
				\includegraphics[width=1\linewidth]{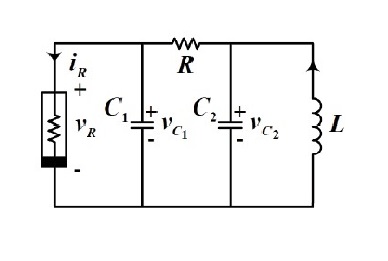}
				\label{figdroopy}
			}
			\quad 
			\subfloat[]
			{
				\includegraphics[width=0.7\linewidth]{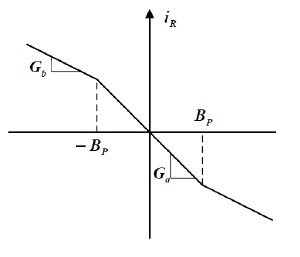}
				\label{figdroopy}
			}
			\caption{The Chua's circuit \citep{aguirre}.
				\\(a) the Chua' circuit and (b) curve of Chua diode (current-voltage characteristic).}
			\label{Figura 1}
		\end{figure}

The Chua's circuit is one of the most used systems as an example to study the non-linear dynamics and chaos. The circuit shown in Figure \ref{Figura 1}a is composed of linear elements, except the Chua's diode, which shows non-linear behaviour as shown in Figure \ref{Figura 1}b. This non-linear element can be implemented by operational amplifiers \citep{aguirre}.The Equations of the circuit are \ref{eq1}, \ref{eq2} and \ref{eq3}.

	\begin{equation}
	C_1 \frac{dv_{c_1}}{dt} =\frac{v_{c_2} - v_{c_1}}{R} - i_d(v_{c_1})
	\label{eq1}
	\end{equation}
	\begin{equation}
	C_2 \frac{dv_{c_2}}{dt} =\frac{v_{c_1} - v_{c_2}}{R} - i_d
	\label{eq2}
	\end{equation}
	\begin{equation}
	L \frac{di_L}{dt} = - v_{c_2} 
	\label{eq3}
	\end{equation}
	
Since $V_{C_1}$ is the voltage across the capacitor $C_1$, $V_{C_2}$ is the voltage across the capacitor $C_2$, $i_L$ is the current through the inductor and the current in the diode is\citep{aguirre}:

\begin{equation}
i_d(v_{C_1})=\left\{ \begin{array}{rcl}
{m_0 v_{C_1} + B_p(m_0 - m_1)} &\mbox{for}
&v_{C_1}< -B_p \\ {m_1 v_{C_1} } &\mbox{for}
&|v_{C_1}| \leq B_p \\
{m_0 v_{C_1} + B_p(m_1 - m_0)} &\mbox{for}
&v_{C_1})> B_p 
\end{array}\right.
\end{equation}

The component values and constants used in the simulations are shown in Table 1. The Algorithms 1 and 2 used in this work are presented in the Appendix. 
		
		\begin{table}[!hbt]
			\caption{Values of components and constants used in the simulations.}
			\centering
			\begin{tabular}{| p{20mm} | c | l |}
				\hline \textbf{Components}& \textbf{Values}\\ \hline
				\centering
				$C_1$& 10 nF\\ \hline
				\centering
				$C_2$& 100 nF\\ \hline
				\centering
				$L$& 19 mH\\ \hline
				\centering
				$R$& 1980 $\Omega$ \\ \hline
				\centering
				$m_0$& -0.37 mS \\ \hline
				\centering
				$m_1$& -0.68 mS\\ \hline
				\centering
				$B_p$& 1.1 V\\ \hline
			\end{tabular}
		\end{table}
		
Algorithm 1 shows the procedure to decrease the error propgation. We use classic RK4, but for each time step, we make two calculations, one using round mode towards $ -\infty $ and other towards $  + \infty $. Algorithm 2 shows that we only make the use of average, that is, we get the mid position for each state variable, from the interval established by the two round modes. This seems quite naive, but it is not. The standard round mode is the round to nearest. In some way, it works perfectly when one is interested in just to get the result for only arithmetic operations or to store a real value. But there is no straight way to make a correct round, when the most important thing is a function, or a complex set of operations. When we make the average, we are in some way approaching the round to the nearest, but from the point of view of functions, what we may call as round to the function. Additionally, as the error coming from rounding off and truncation can be se as random (or at least pseudo-random), the average can be also seen as a filter to reduce the total error.

	\section{RESULTS}

The results of Chua's circuit simulation according to the values in Table 1 and using the IEEE 754-2008 standard for floating point are shown in Figures 2a and 3a. In these first results we use an traditional approach by means of RK4.  However, the results from the solution proposed by Algorithm 1, but using the same parameters and initial condition, as shown in Table 1, are presented in Figures 2b and 3b.
	\begin{figure}[!htb]
	\centering
	\subfloat[]
	{
		\includegraphics[width=0.95\linewidth]{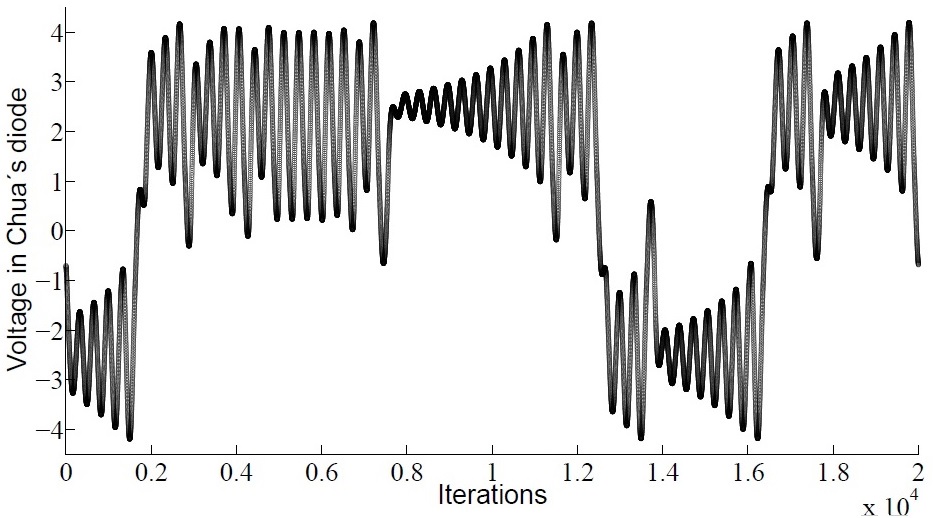}
		\label{figsnoop}
	}
	
	\quad 
	\subfloat[]
	{
		\includegraphics[width=0.95\linewidth]{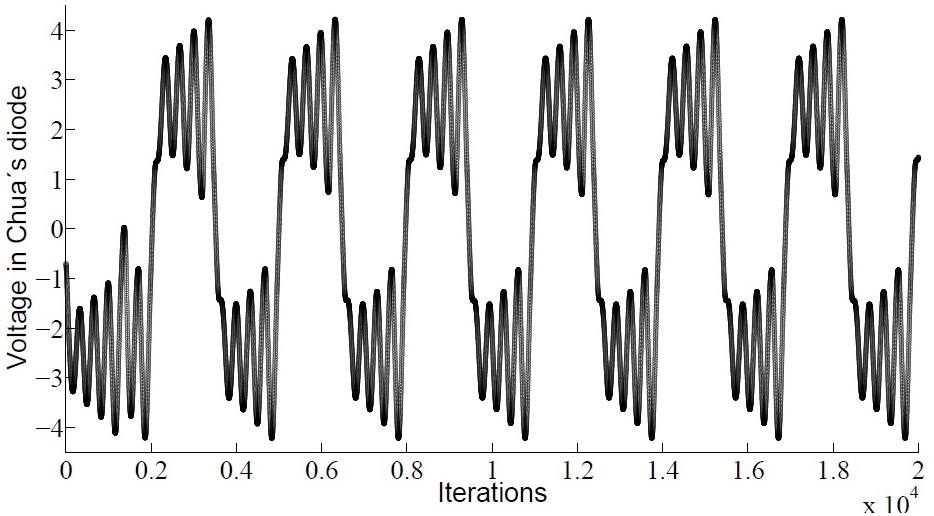}
		\label{figsnoop}
	}
		
		\begin{center}
			\caption{Voltage in Chua's diode.
				\\ (a) Voltage in the Chua's diode when the simulation was traditionally performed using standard IEEE 754-2008. (b) Voltage in the Chua's diode when the simulation is performed by the average of the two round modes towards to $ -\infty $ and $ +\infty $.} 			
		\end{center}	
	\end{figure}
	
	\graphicspath{{figuras/}{fig_site/}}
	
	\begin{figure}[!htb]
		\centering
		\subfloat[]
		{
			\includegraphics[width=1.1\linewidth]{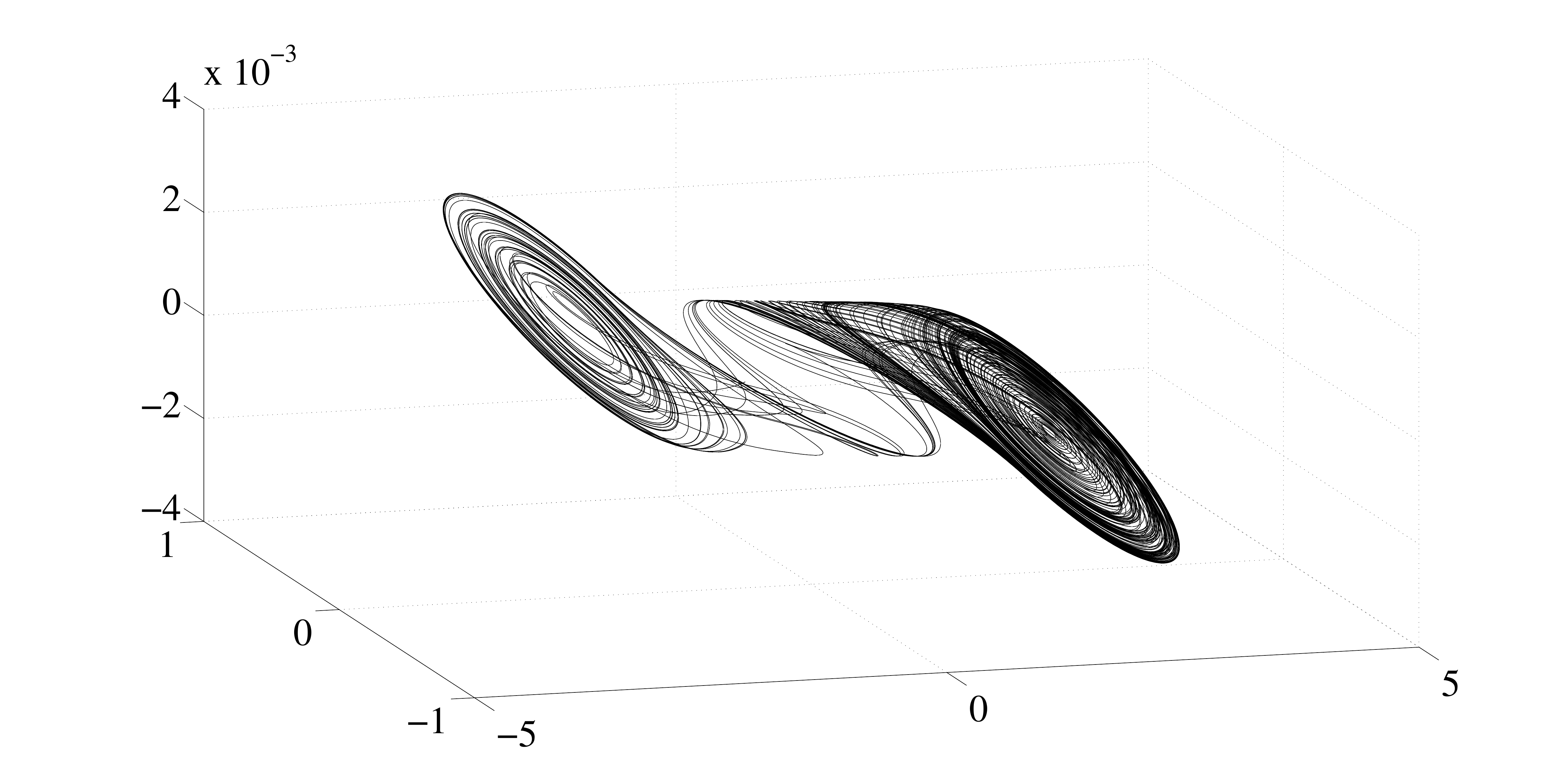}
			\label{figsnoop}
		}
		
		\quad 
		\subfloat[]
		{
			\includegraphics[width=1.1\linewidth]{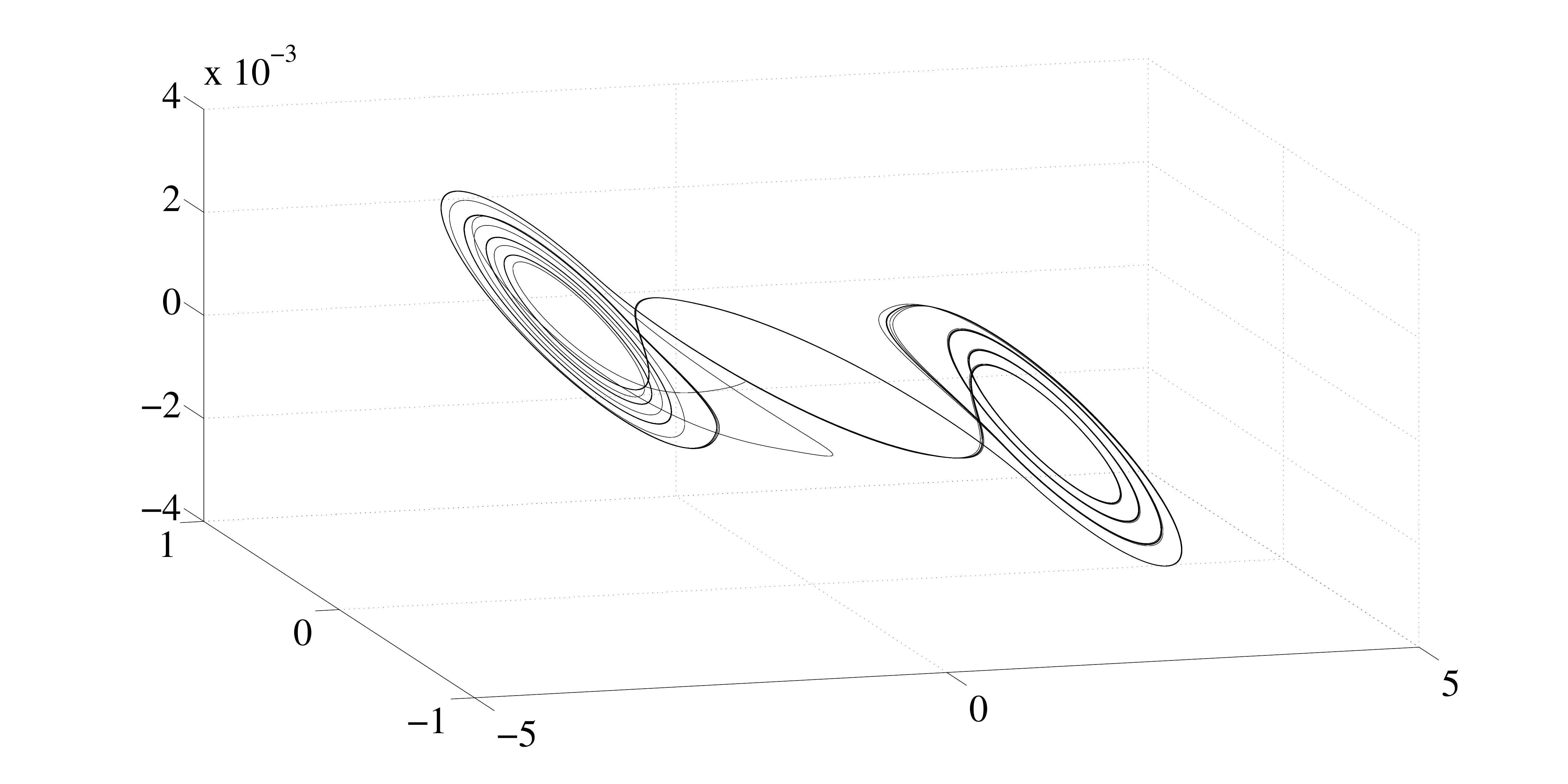}
			\label{figsnoop}
		}
		\begin{center}
			\caption{Scroll of Chua's circuit.
				\\ (a) Double attractor traditionally generated. (b) Periodical result by the proposed algorithm.} 
			
		\end{center}	
	\end{figure}
It is known that the diode curve, Figure 1b, is given by the characteristic voltage \textit{versus} current and  in Figure 2 has the diode voltage. Thus, it is found by applying the intervals analysis Chua's circuit, the voltage begins to show a periodic behavior, Figure 2b. Thus, the diode curve changes and thus the dynamics of the system is affected as shown in Figure 3b.

The orbit of a chaotic dynamic system is sensitive to initial conditions and aperiodic \citep {chaos}. Although, for the same set of parameters and initial conditions, Figure 2b shows a periodical result, which is not sensitivity to initial condition.

\section{CONCLUSION}
It is observed that even though infinitesimal errors along the simulation can compromise the result. Interval analysis were employed as the round mode was adapted to set the mid value of an interval composed by the round modes towards to $ -\infty $ and $ +\infty $. So, what it could be noticed is that the system simulated by tradition RK4 and IEEE-754-2008 round mode standard, that is, round to the nearest, presents a strange attractor with a probably chaotic behaviour. On the contrary, the proposed algorithm furnishes a beautiful periodical result. Which one is the correct? We do not know for sure. The simplicity of a periodical solution and the reasons presented to use an average of the round methods, help us to believe that this is the correct solution.  We intend to build Chua's circuit following the steps of \citep{kennedy92} and compare this simulated results with experimental results. Although, this procedure seems a way to give a final answer, we should keep in mind that there is no way to implement a circuit that matches perfectly the set of equations and parameters given in this paper. This observation makes our investigation yet more fascinating!

 \bibliographystyle{apalike}
 \bibliography{exemplo}

\begin{thebibliography}{}

\bibitem[Aguirre, 2007]{aguirre}
Aguirre, L.~A. (2007).
\newblock {\em Introdu\c{c}\~{a}o \`{a} Identifica\c{c}\~{a}o de Sistemas -
  T\'{e}cnicas Lineares e N\~{a}o-Lineares Aplicadas a Sistemas Reais}.
\newblock Editora da UFMG.
\newblock $3^a$ edi\c{c}\~{a}o.

\bibitem[Chua, 1994]{chua94}
Chua, L.~O. (1994).
\newblock Chua's circuit: Ten years later. 11.
\newblock {\em IEEE Trans. Fundamentals.}, E-77A:1811--1822.

\bibitem[Goldberg, 1991]{Gold}
Goldberg, D. (1991).
\newblock What every computer scientist should know about floating-point
  arithmetic.
\newblock {\em Computing Surveys}, 23:5–48.

\bibitem[{Institute of Electrical and Electronics Engineers (IEEE)},
  2008]{ieee}
{Institute of Electrical and Electronics Engineers (IEEE)} (2008).
\newblock {\em {IEEE} 754 standard for floating-point arithmetic}.

\bibitem[Kennedy, 1992]{kennedy92}
Kennedy, M.~P. (1992).
\newblock Robust op amp realizaion of chua's circuit.
\newblock {\em Frequenz}, 46(3-4):66--80.

\bibitem[Monteiro, 2002]{Monteiro}
Monteiro, L. H.~A. (2002).
\newblock {\em Sistemas Din\^{a}micos}.
\newblock Editora Livraria da F\'{i}sica.

\bibitem[Moore et~al., 1979]{moore1979methods}
Moore, R.~E., Bierbaum, F., and Schwiertz, K.-P. (1979).
\newblock {\em Methods and applications of interval analysis}, volume~2.
\newblock SIAM.

\bibitem[Nepomuceno, 2014]{Nepomuceno2014}
Nepomuceno, E.~G. (2014).
\newblock Convergence of recursive functions on computers.
\newblock {\em The Journal of Engineering}, pages 1--3.

\bibitem[S.~T.~Alligood, 1997]{chaos}
S.~T.~Alligood, T. D.~Sauer, J. A.~Y. (1997).
\newblock {\em Chaos: An Introduction to Dynamical Systems}.
\newblock Springer.

\end{thebibliography}

\lstset{language=matlab}          

\newpage
\section{Appendix}

\subsection{Algorithm 1}	
\begin{lstlisting} 
%Initial Conditions
clear all
system_dependent('setround',-Inf);
ym=[-0.7 0 0];
system_dependent('setround',Inf);
yp=[-0.7 0 0];
tf=0.075;
h=1e-6;
tspan = 0:h:tf;
N=length(tspan);
for k=1:N-1
  system_dependent('setround',-Inf);
  aux = ode4(@chua,tspan(k:k+1),
  ym(k,:),yp(k,:));
  ym(k+1,:)=aux(2,:);
  system_dependent('setround',Inf);
  aux = ode4(@chua,tspan(k:k+1),
  yp(k,:),ym(k,:));
  yp(k+1,:)=aux(2,:);
end
%Figures
figure(1)
plot(1:N,ym(:,1),1:N,yp(:,1),'k')
figure(2)
plot3(ym(:,1),ym(:,2),ym(:,3),'k')
view(-16,24);grid;
\end{lstlisting}

\subsection{Algorithm 2}	
\begin{lstlisting} 			
function out = chua(t,in,in2)
  x = in(1); y = in(2); z = in(3); 
  x2=in2(1);y2=in2(2);z2=in2(3);
  L = 19.2*10^(-3);
  C1  = 10*10^(-9);
  C2  = 100*10^(-9);
  R = 1978.5;      
  G = 1/R;
  m0=-0.37*10^(-3);
  m1=-0.68*10^(-3);
  Bp=1.1;
  %Average +Inf and -Inf			
  x=(x+x2)/2;
  y=(y+y2)/2;
  z=(z+z2)/2;	
  %Diode
  if  x >Bp
    g=m0*x+Bp*(m1-m0); 
  elseif  (x >= -Bp)&(x <= Bp)
    g=m1*x;
  else
    g=m0*x+Bp*(m0-m1); 
  end
  % Chua's Circuit Equations
  xdot = (1/C1)*(G*(y-x)-g);
  ydot = (1/C2)*(G*(x-y)+z);
  zdot  = -(1/L)*y;
  out = [xdot ydot zdot]';
\end{lstlisting}
\end{document}